\documentclass[%
 reprint,
superscriptaddress,
 amsmath,amssymb,
 aps,
prb,
]{revtex4-1}

\usepackage{color}
\usepackage{graphicx}
\usepackage{breqn}
\usepackage{dcolumn}
\usepackage{bm}

\usepackage{xspace}

\newcommand{\ket}[1]{|#1\rangle}
\newcommand{\bra}[1]{\langle#1|}
\newcommand{\braket}[2]{\langle#1|#2\rangle}
\newcommand{\expect}[1]{\langle #1 \rangle}

\begin{document}

\preprint{APS/123-QED}

\title{Symmetry Breaking in Occupation Number Based Slave-Particle Methods}

\author{Alexandru B. Georgescu}
\affiliation{Department of Physics}
\affiliation{Center of Research on Interface Structure and Phenomena}
\author{Sohrab Ismail-Beigi}
\affiliation{Department of Applied Physics}
\affiliation{Department of Physics}
\affiliation{Center of Research on Interface Structure and Phenomena}
\affiliation{Department of  Mechanical Engineering and Materials Science\\Yale University$,$ New Haven$,$ CT 06520$,$ USA}




\date{\today}

\begin{abstract}

We describe a theoretical approach for finding spontaneously symmetry-broken electronic phases due to strong electronic interactions when using recently developed slave-particle (slave-boson) approaches based on occupation numbers.  We describe why, to date, spontaneous symmetry breaking has proven difficult to achieve in such approaches.  We then provide a total-energy based approach for introducing auxiliary symmetry breaking fields into the solution of the slave-particle problem that leads to lowered total energies for symmetry broken phases.  We point out that not all slave-particle approaches yield to energy lowering: the slave-particle model being used must explicitly describe the degrees of freedom that break symmetry.  Finally, our total energy approach permits us to greatly simplify the formalism used to achieve a self-consistent solution between spinon and slave modes while increasing numerical stability and greatly speeding up the  calculations.

%
\end{abstract}

\maketitle

\section{Introduction}

The effects of strong electronic interactions and electronic correlations on materials properties is a subject with a considerable history.  The most celebrated textbook example is the Mott transition where by increasing the strength of localized electronic repulsions, the electrons in the material lose band mobility and instead localize on the atomic sites (i.e., loss of wave behavior).  However, electronic correlations also underlie many other ordered electronic phases such as various forms of magnetism as well as superconductivity.  A canonical model Hamiltonian for correlated electron is the (extended) Hubbard model where electrons can hop between localized orbitals centered on atomic sites but multiple electronic occupancy of a given atomic site leads to a significant energy penalty $U$.  By varying the ratio of $U$ to the band hopping parameters, one can cover the range from weak to strong electronic interactions and correlations.\cite{HubbardModel}

The workhorse in realistic first principles calculations in crystal and electronic structure calculations, Density Functional Theory (DFT)\citep{HK},
is fundamentally based on a description of non-interacting electrons, i.e., band theory.  Due to its simple structure, band theory approaches can not capture the effects of dynamical electronic fluctuations and localized correlations on  electronic band spectra.  Extensions of DFT to go beyond local exchange-correlation potentials and to include non-local Hartree-Fock type electronic behavior, such as the DFT+U or hybrid functional approaches\citep{Becke1993,Perdew1996}, can capture certain effects of electron-electron interactions especially for strongly symmetry-broken situations.  Nevertheless, these are still band theory descriptions incapable of leading, e.g., to electron localization without resorting to symmetry breaking.  

More advanced computational many-body approaches for simulation of electronic correlations are based on Green's functions methods.  One type of approach is the $GW$ approximation to the electron self-energy \cite{hedin_new_1965,hybertsen_electron_1986,aryasetiawan_gw_1998} which is a fully {\it ab initio} approach that includes the physics of non-local and dynamical electronic screening and produces accurate results for electronic band energies of a wide variety of materials \cite{aryasetiawan_gw_1998,van_schilfgaarde_quasiparticle_2006}.  However, the $GW$ method is based on summation of a subset of many-body diagrams (RPA diagrams) and thus does not capture a number of physical effects; separately $GW$ calculations are notoriously expensive in terms of computation time due to their fully {\it ab initio} nature and lack of a particular basis set. Another avenue of approach is represented by Dynamical Mean Field Theory (DMFT) \cite{Georges1996,Kotliar2006} which can include the effect of local interactions and dynamical fluctuations by solving a model Hamiltonian with local interactions exactly (i.e., all diagrams for the local interactions are included). However, DFT+DMFT calculations on realistic materials with large unit cells are still quite challenging as they require large-scale parallel computations.

For all these reasons, approximate and efficient methods for solving correlated problems continue to be of interest to the computational many-body community.  One set of methods of recent interest for solving Hubbard models are slave-particle (slave-boson) methods. This  method that has a long background in condensed matter theory. These method have been used to study cases with infinitely strong repulsive interactions.  \cite{Barnes1976,Barnes1977,Coleman1983,Read1983,Read1985,Kotliar1986,Lee2006}   Dealing with finite interaction strengths was enabled by Kotliar-Ruckenstein approach\citep{Kotliar1986} whose variants and modifications have been applied to study high-temperature superconductors \citep{Raczkowski2006} as well as  multi-band models \citep{Fresard1996,Lechermann2007,Bunemann2011} to elucidate the effects of multiple orbitals, degeneracy and Hund's coupling. \citep{Fresard1996,Lechermann2007}  In these approaches, each bosonic slave degree of freedom tracks the occupancy of a particular electronic configuration of a correlated site: once multiple orbitals and multiple electron counts can exist on a site, the number of require bosons becomes large.  These methods can and have been used to describe spontaneously broken electronic symmetry (e.g., magnetic) states.\cite{Kotliar1986,fresard_slave-boson_1991,Fresard1996} 

A recent set of more economical slave-particle methods has been developed and have become of wider interest, such as the slave-rotor method \citep{Florens2002, Florens2004} and its application to nickelate oxides \citep{Lau2013a} and the slave-spin method \citep{DeMedici2005,Hassan2010} and its application to iron-based superconductors \cite{DeMedici2014}. Recently, we have developed a generalized version of these methods that does not require the analogy with spin or angular momentum and introduces multiple intermediate slave-particle models.\cite{Georgescu2015} These recent approaches use slave degrees of freedom to track the electron occupation number on a site, and its distribution among orbital and spin channels, and thus require a much smaller number of bosons per site. 

However, in all the previous literature in which these occupation number based methods has been used, spontaneous symmetry breaking has been achieved in multi-orbital systems where both a Hubbard $U$ as well as a non-zero Hund's $J$ interaction have been operative.\cite{Hassan2010,DeMedici2014,Lau2013} For a system where only the repulsion $U$ operates, spontaneous symmetry breaking has not been displayed even when interaction-induced magnetism is a feature of the actual ground state of the model Hamiltonian (e.g., ground-state antiferromagnetic order for a half-filled single-band Hubbard model). Indeed, as we show, stabilizing a purely interaction induced symmetry-broken phase is very difficult for slave-particle methods without introduction of symmetry breaking fields. Our work describes this issue in  detail and provides a total-energy approach that naturally produces symmetry breaking.  We then show how one can make slave-particle self-consistency between spinon and slave modes much more efficient via a specific and exact decoupling of the two modes.

\section{The Slave-Particle Approach}

In this section we review the key aspects of the slave-particle formalism used in previous work to set up the notation and language used in subsequent sections.  The general correlated-electron Hamiltonian we consider is an extended Hubbard model given by
\begin{multline}
\hat{H}=\sum_i \hat{H}_{int}^i +\sum_{im\sigma}\epsilon_{im\sigma}\hat{d}^\dag_{im\sigma}\hat{d}_{im\sigma}\\
-\sum_{ii'mm'\sigma} t_{imi'm'\sigma}  \hat d^\dag_{im\sigma} \hat d_{i'm'\sigma}\,.
\label{eq:HHubbard}
\end{multline}
The $\hat d$ are canonical fermion annihilation operators.  The indices $i,i'$ range over the localized sites in the system (usually atomic sites), $m,m'$ range over the localized spatial orbitals on each site, $\sigma=\pm1$ denotes spin, $\hat{H}_{int}^i$ is the local Coulombic interaction for site $i$, $\epsilon_{im\sigma}$ is the onsite energy of the state labeled by $im\sigma$, and $t_{imi'm'\sigma}$ is the spin-conserving hopping element term connecting orbital $im\sigma$ to $i'm'\sigma$.  A commonly used interaction term is given by the Slater-Kanamori form~\cite{Kanamori1963}
\begin{multline}
\hat{H}_{int}^i=\frac{U_i}{2}(\hat n_i^2-\hat n_i) + \frac{U_i-U'_i}{2}\sum_{m\ne m'}\hat n_{im}\hat n_{im'} \\
- \frac{J_i}{2} \sum_\sigma\sum_{m\ne m'}\hat n_{im\sigma}\hat n_{im'\sigma}\\
- \frac{J_i}{2}\sum_\sigma \sum_{m\ne m'}\left( 
\hat d^\dag_{im\sigma}\hat d_{im\bar\sigma}\hat d^\dag_{im'\bar\sigma}\hat d_{im'\sigma} \right.\\
\left. + \hat d^\dag_{im\sigma}\hat d^\dag_{im\bar\sigma}\hat d_{im'\sigma}\hat d_{im'\bar\sigma}
\right)
\end{multline}
While the Coulombic parameters $U_i$, $U'_i$ and $J_i$ can in principle depends the site index $i$, in practice in most models they are assumed to be the same for all correlated sites.  Briefly, the $U$ term describes repulsion between the same spatial orbitals on a site, $U'$ repulsion between different orbitals, and $J$ measures the strength of the Hund's interaction between different orbitals with the same spin state.  The number operators are
\[
\hat n_{im\sigma} = \hat d^\dag _{im\sigma}\hat d_{im\sigma} \ , \ \hat n_{im} = \sum_\sigma \hat n_{im\sigma} \ , \ \hat n_i = \sum_{m\sigma} \hat n_{im\sigma}\,.
\]

The interacting Hubbard problem is impossible to solve exactly and even difficult to solve approximately.  Some of the complexity is due to the fact that the interacting fermions have both charge and spin degrees of freedom. In slave-boson approaches\cite{Barnes1976,Barnes1977,Coleman1983,Read1983,Read1985,Kotliar1986,Lee2006}, one 
separates the spin from charge degrees of freedom at each site by introducing a spinless charged bosonic ``slave'' degree of freedom on each site along with a spinful neutral fermion termed a spinon.  The spinon and slave boson annihilation operators are indicated by $\hat f$ and $\hat O$ operators, respectively.   Specifically, the electron field operators is decomposed as
\begin{equation}
\hat{d}_{im\sigma}=\hat{f}_{im\sigma}\hat{O}_{i\alpha} \ , \
\hat{d}_{im\sigma}^{\dagger}=\hat{f}_{im\sigma}^{\dagger}\hat{O}_{i\alpha}^{\dagger}\,.
\label{eq:slaveopdef}
\end{equation}
The index $\alpha$ is part of our generalized notation~\cite{Georgescu2015}  that permits us to unify different occupation number based slave-particle models.  The meaning of $\alpha$ depends on the type of slave boson model chosen, and $\alpha$ refers to a subset of the $m\sigma$ indices that belong to a site $i$.  For example, if we use a slave-rotor model for the correlated orbitals on a site \citep{Florens2002, Florens2004}, then $\alpha$ is nil: $\hat O_{i\alpha}=\hat O_i$. Namely, we have a single slave particle on each site $i$ that only tracks the total number of electrons on that site.  At the opposite limit, we can have a unique slave boson for each $m\sigma$ combination on a site (the ``slave-spin'' method\citep{DeMedici2005, Hassan2010}), so that in this case $\alpha=m\sigma$.

The introduction of slave bosons by itself does not make solution of the Hubbard model any easier as more degrees of freedom have been introduced to further enlarge the Hilbert space.  To avoid sampling of unphysical states in the enlarged spinon+slave Hilbert space which have no correspondence to in the original electronic Hilbert space, one must ensure that the number of slave particles and number of spinons track each other. More precisely, Eq.~(\ref{eq:slaveopdef}) shows, spinon and slave particles are created or annihilated at the same time so that only state kets in the extended Hilbert space that obey this condition are physical.  Hence, one must ensure that
\[
\hat d^\dag_{im\sigma} \hat d_{im\sigma} = \hat f^\dag_{im\sigma} \hat f_{im\sigma}
\]
and also that the subset of physical states $\ket{\Psi_{phys}}$ must obey 
\begin{dmath}
\hat n_{i\alpha}\ket{\Psi_{phys}}=\hat{N}_{i\alpha}\ket{\Psi_{phys}}
\label{eq:opconstraint}
\end{dmath}
where $\hat N_{i\alpha}$ is the number counting operator for the slave particles and the correspond particle count for spinons is
\begin{equation}
\hat n_{i\alpha} = \sum_{m\sigma\in\alpha}\hat{f}_{im\sigma}^\dag\hat{f}_{im\sigma}\,.
\end{equation}
This constraint on the physical states simply ensures that the number of slave bosons matches exactly the number of spinons on each site.

The key approximation that makes the slave-boson approach more tractable than the original problem is to assume a separable form for the overall wave function of the system which takes a product form $\ket{\Psi_f}\ket{\Phi_s}$ where $\ket{\Psi_f}$ is a spinor-only state ket and $\ket{\Phi_f}$ is a slave-only state ket.  This means one can only enforce the above operator constraints on average:
\begin{equation}
\expect{\hat n_{i\alpha}}_f = \expect{\hat{N}_{i\alpha}}_s
\label{eq:avgmatch}
\end{equation}
where the spinon and slave averages for any operator $\hat A$ are defined via
\[
\expect{\hat A}_f = \bra{\Psi_f}\hat A\ket{\Psi_f} \ , \ 
\expect{\hat A}_s = \bra{\Phi_s}\hat A\ket{\Phi_s} \,.
\]
This separability assumption means one must solve two separate and easier eigenvalue problems
\[
\hat H_f \ket{\Psi_f} = E_f \ket{\Psi_f} \ , \ \hat H_s\ket{\Phi_s} = E_s \ket{\Phi_s}
\]
in a self-consistent fashion.  The spinon Hamiltonian is given by
\begin{multline}
\hat{H}_f=\sum_{im\sigma}\epsilon_{im\sigma}\hat{f}_{im\sigma}^\dag\hat{f}_{im\sigma}
-\sum_{i\alpha}h_{i\alpha}\hat n_{i\alpha}\\
-\sum_{ii'\alpha\alpha'} \langle\hat{O}^\dag_{i\alpha}\hat{O}_{i'\alpha'}\rangle_s\!
\sum_{\substack{m\sigma\in\alpha\\m'\sigma\in\alpha'}}
t_{imi'm'\sigma}  \hat{f}^\dag_{im\sigma}\hat{f}_{i'm'\sigma}\,.
\label{eq:Hspinon}
\end{multline}
The slave boson Hamiltonian takes the form
\begin{multline}
\hat{H}_{s}=\sum_i \hat H^i_{int} +\sum_{\alpha}h_{i\alpha}\hat{N}_{i\alpha}\\
-\sum_{ii'\alpha\alpha'}\left[
\sum_{\substack{m\sigma\in\alpha\\m'\sigma\in\alpha'}}t_{imi'm'\sigma}\langle \hat{f}^\dag_{im\sigma}\hat{f}_{i'm'\sigma} \rangle_f \right]
 \hat{O}^\dag_{i\alpha}\hat{O}_{i'\alpha'}
 \label{eq:Hslave}
\end{multline}
where  the spinon averages $\langle \hat{f}^\dag_{im\sigma}\hat{f}_{i'm'\sigma} \rangle_f$ renormalize the slave boson hoppings.  The slave boson problem is one of interacting charged bosons without spin on a lattice.

Self-consistency refers to the fact that the spinon Hamiltonian involves averaged quantities involving the slave wave function and vice versa.  In addition, the values of the Lagrange multipliers $h_{i\alpha}$ must be chosen to ensure average particle number matching as per Eq.~(\ref{eq:avgmatch}).

\section{Single-site mean-field approximation}
\label{sec:meanfield}

In practice, the slave Hamiltonian of Eq.~(\ref{eq:Hslave}) represents a many-body interaction bosonic problem that has no exact solution.  In what follows, when solving numerically for the ground state of a spinon+slave problem, we will use a simple single-site mean-field approach: when dealing with site $i$ in the salve problem, we replace the $\hat O_{i\alpha}$ slave operators on the other neighboring sites by their averages $\langle \hat O_{i\alpha}\rangle_s$.  For the spinon Hamiltonian, this boils down to the simple replacement
\[
\langle\hat{O}^\dag_{i\alpha}\hat{O}_{i'\alpha'}\rangle_s \rightarrow \langle\hat{O}^\dag_{i\alpha}\rangle_s \langle \hat{O}_{i'\alpha'}\rangle_s
\]
in the hopping term.  The slave Hamiltonian turns into
\begin{multline}
\hat{H}_{s}=\sum_i \hat H^i_{int} +\sum_{\alpha}h_{i\alpha}\hat{N}_{i\alpha}\\
-\sum_{ii'\alpha\alpha'}\left[
\sum_{\substack{m\sigma\in\alpha\\m'\sigma\in\alpha'}}t_{imi'm'\sigma}\langle \hat{f}^\dag_{im\sigma}\hat{f}_{i'm'\sigma} \rangle_f \right]\cdot \\
\left[ \langle \hat{O}^\dag_{i\alpha}\rangle_s \hat{O}_{i'\alpha'} + h.c. \right]
\end{multline}
which is a simple many-body system of isolated  sites where the bosonic $\hat O_{i\alpha}$ and $\hat O_{i\alpha}^\dag$ operators remove and add bosons to the site from an effective bosonic mean-field bath.  We note that, for the simple model Hamiltonians we will be using below in this approach, the quasiparticle renormalization factor (or weight) $Z$ is simply given by $Z_{i\alpha} = \langle O_{i\alpha}\rangle_s^2$.

\section{Difficulties Obtaining Symmetry Broken Phases}

In this section, we explain why the current implementation of mean-field theory fails to obtain proper symmetry broken phases. We use the example of the well-understood one-dimensional Hubbard model at half filling.  Consider the Hamiltonian:
\begin{equation}
\hat{H}=\frac{U}{2}\sum_i(\hat{N}^2_i-\hat{N}_i)-\sum_{i,\sigma}t(\hat{c}^{\dag}_{i,\sigma}\hat{c}_{i+1,\sigma}+\hat{c}^{\dag}_{i+1,\sigma}\hat{c}_{i,\sigma})
\label{eq:1dHubH}
\end{equation}
where $i$ is the site index, there is a single orbital per site, there are two spin channels per site, and we consider the case where we are at half filling ($\langle \hat {N}_i\rangle=1$).
The ground state is well-known. For $U=0$, the ground state is non-magnetic and metallic. For $U>0$ but finite, the ground state is insulating and shows anti-ferromagnetic correlations \citep{Lieba2009} but has finite quasiparticle weight $Z>0$.  

The $U=0$ and $U>>|t|$, the model's solutions are well-described by existing slave-particle mean-field implementations. For the intermediate region $U\sim|t|$, we are aware of no published study using recent slave-spin, slave-rotor or other formalisms from the same family that has correctly obtained the correct AFM phase for this model.  Namely, the AFM solution does not appear to be a self-consistent ground state solution of the spinon+slave coupled Hamiltonians.  In addition to being annoying, this is very worrisome since even a simple uncorrelated approach such as Hartree-Fock easily delivers an AFM ground state.

To understand where the problem lies, consider the spinon Hamiltonian of Eq.~(\ref{eq:Hspinon}) and how one would achieve symmetry breaking, e.g., spin symmetry breaking and ordering, due to electron interaction effects.  Since the electron interaction is handled by the slave sector, the only quantities that can be affected by the slave calculation that then feed into the spinon Hamiltonian are the Lagrange multiplies $h_{i\alpha}$ and the  rescaling factors $\langle\hat{O}^\dag_{i\alpha}\hat{O}_{i'\alpha'}\rangle_s$ of the spinon hopping.  

In the simplest slave treatment, we have a single slave particle on the site: for example, the slave-number or slave-rotor treatments.  In such a case, the $\alpha$ label is nil so our Lagrange multipliers are only indexed by site $h_i$ and the rescaling factors as well $\langle\hat{O}^\dag_{i}\hat{O}_{i'}\rangle_s$.  Obviously, no spin symmetry breaking is possible in the spinon sector since these variables do not depend on spin in any way.

When we move to more elaborate slave-particle models where there are different slave modes for the different spin channels, then one can imagine that symmetry breaking is possible.  For example, in our single orbital per site 1D Hubbard model, when we have one slave-particle for each spin channel, then $\alpha=\sigma$.  We could now imagine that   the $h_{i\sigma}$ shift the on-site energies of the orbitals in such a way to break spin symmetry, or that the hopping rescaling factors are also spin dependent.  In practice, however, we have not found this to be the case: starting from a strongly symmetry broken initial guess, the self-consistency cycle between spinon and slave sectors drives the system towards a paramagnetic solution and the two spin channels become equivalent.  Any initial magnetization disappears upon self-consistent iteration.

\begin{figure}
\includegraphics[width=3in]{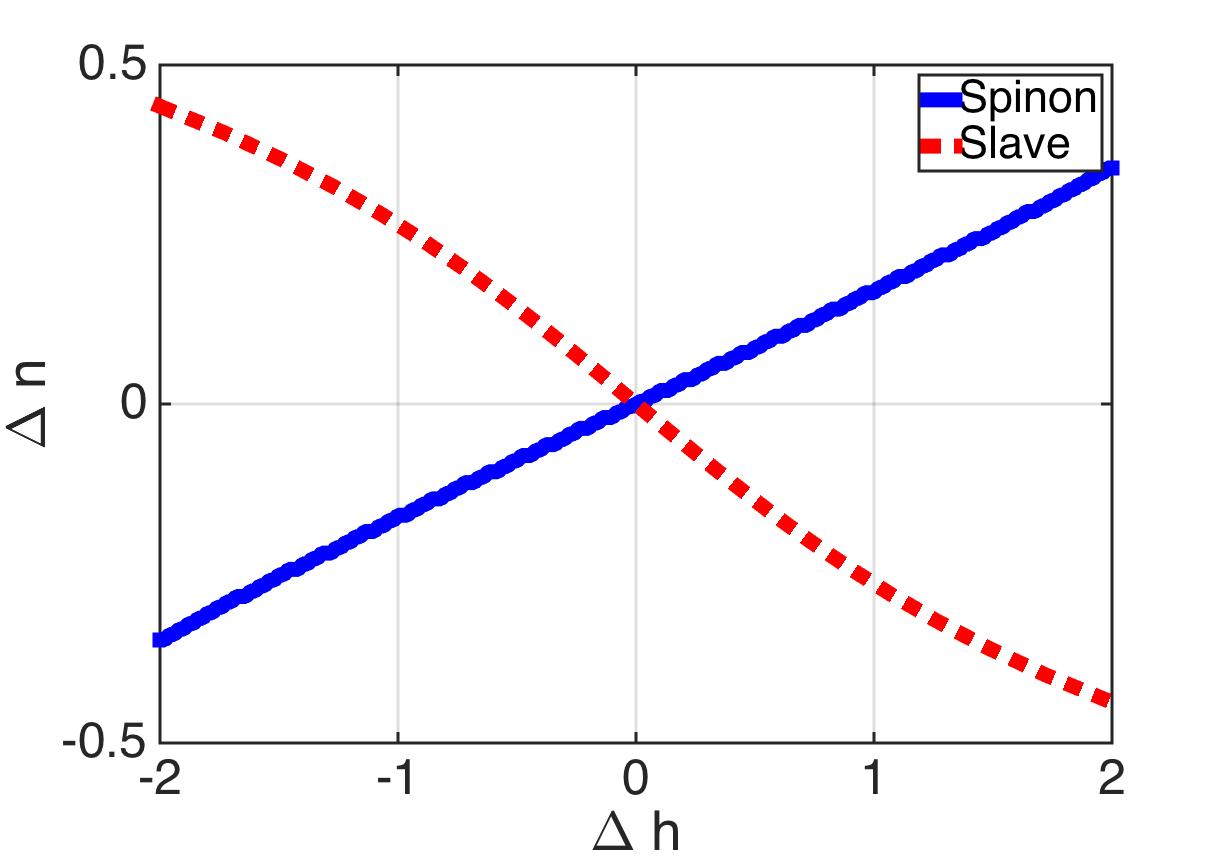}
\includegraphics[width=3in]{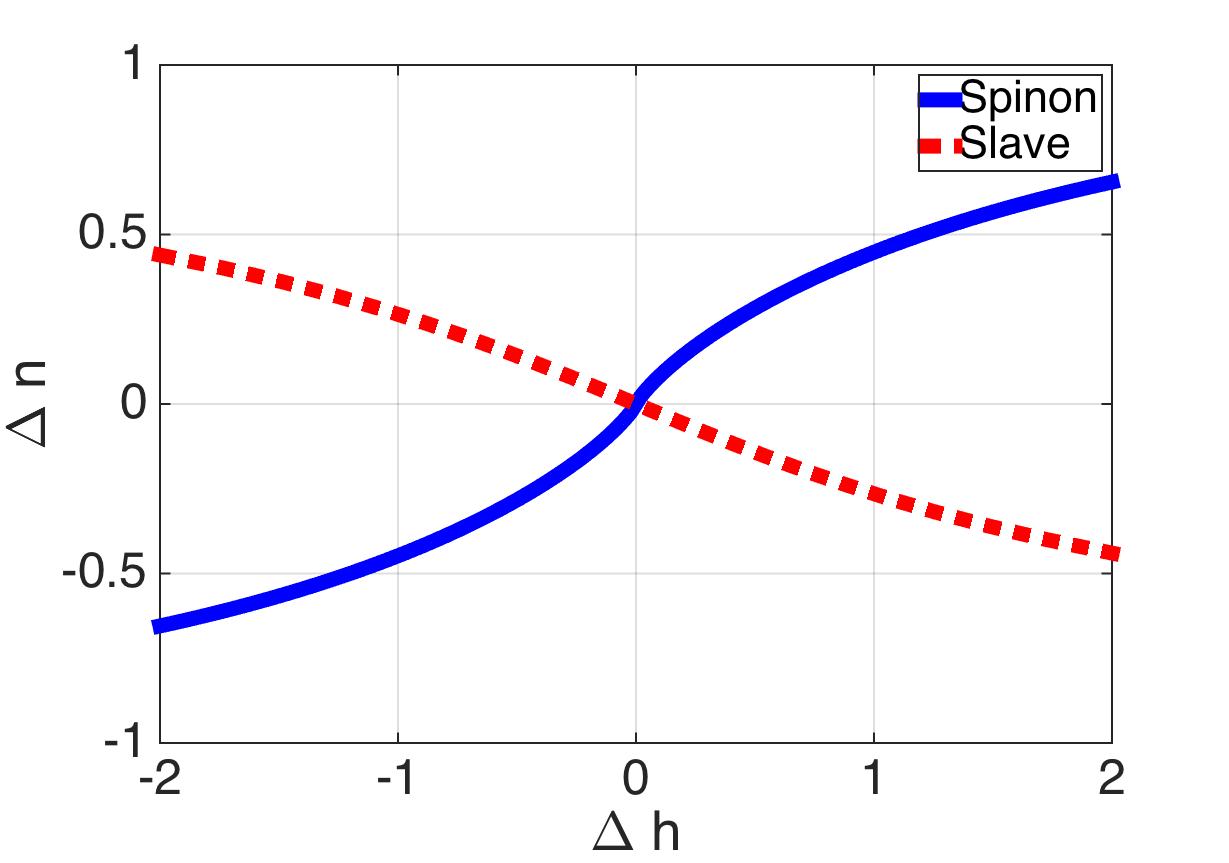}
\caption{\label{fig:symmproof} $\Delta n=n_{\uparrow}-n_{\downarrow}$ as a function of $\Delta h=h_{\uparrow}-h_{\downarrow}$ on one site of the 1D half-filled single band Hubbard model with $U=2$ and $t=1$.  Upper figure is for the FM phase, and the lower figure for the AFM phase.  The $\Delta h$ dependence of the spinon and slave occupancies are shown separately.  Self-consistency between the two requires zero occupancy difference.
}
\end{figure}

We have analyzed this failure and discovered the following situation.  If at some point the spinon system has broken spin symmetry on a site $i$ with net spin up, then $h_{i\uparrow}> h_{i\downarrow}$ is what makes this true.  However, although $h_{i\uparrow}> h_{i\downarrow}$ favors higher spin $\uparrow$ occupancy in the spinon sector (due to the negative sign in front of $h_{i\alpha}$ in Eq.~(\ref{eq:Hspinon})), it favors higher occupancy of the spin $\downarrow$ channel in the slave sector (positive sign of $h_{i\alpha}$ in Eq.~(\ref{eq:Hslave})).  The two effects fight each other, and the final self-consistent solution has $h_{i\uparrow}=h_{i\downarrow}$.  An explicit example is provided by the 1D single-band Hubbard model at half filling where the dependence of slave and spinon occupancies on $h$ are shown in Figure~\ref{fig:symmproof}.  These plots are generated by providing $\Delta n_i=n_{i\uparrow}-n_{i\downarrow}$ on some fixed site $i$ as input to the slave problem which yields $\Delta h_{i}=h_{i\uparrow}-h_{i\downarrow}$ and $\langle \hat O_{i\sigma} \rangle$ which are then used to solve the spinon problem to get the spinon $\Delta n_i$.  The figures clearly show that the only self-consistent solution where slave and spinon particle numbers match is for $\Delta h_i=0$ which is the symmetric paramagnetic state.

\section{Symmetry breaking fields}
In this section, we show how manually adding small external symmetry breaking terms (``fields'') to the on-site energies can lead to electronic symmetry breaking and lower the energy of the self-consistent ground state. In the next section, we will justify this apparently {\it ad hoc} approach.

Adding additional symmetry breaking (``magnetic field'') terms $b_{im\sigma}$ to the on-site energies of the orbitals in the spinon Hamiltonian gives the simple modification
\begin{multline}
\hat{H}_f=\sum_{im\sigma}\epsilon_{im\sigma}\hat{f}_{im\sigma}^\dag\hat{f}_{im\sigma}
-\sum_{i\alpha}h_{i\alpha}\sum_{m\sigma\in\alpha}\hat{f}_{im\sigma}^\dag\hat{f}_{im\sigma}\\
-\sum_{ii'\alpha\alpha'} \langle\hat{O}^\dag_{i\alpha}\hat{O}_{i'\alpha'}\rangle_s\!
\sum_{\substack{m\sigma\in\alpha\\m'\sigma\in\alpha'}}
t_{imi'm'\sigma}  \hat{f}^\dag_{im\sigma}\hat{f}_{i'm'\sigma}\\
-\sum_{im\sigma}b_{im\sigma}\hat{f}_{im\sigma}^\dag\hat{f}_{im\sigma}\,.
\label{eq:Hspinonwithfield}
\end{multline}
We do not modify the slave Hamiltonian in any way in this {\it ad hoc} approach.

Addition of non-zero symmetry breaking fields $b_{im\sigma}$ will modify the self-consistent solution to the spinon+slave problem.  To gauge if this improves the solution, we monitor the total electronic energy and see if it is lowered due to symmetry breaking.  The total energy is the expectation value of the original Hubbard Hamiltonian of Eq.~(\ref{eq:HHubbard}) with respect to the approximate spinon+slave wave function $\ket{\Psi_f}\ket{\Phi_s}$, and is equal to
\begin{multline}
E_{total}= \langle \hat H \rangle = \sum_i \langle \hat H^i_{int}\rangle_s+\sum_{im\sigma}\epsilon_{im\sigma} \langle\hat {f}^\dag_{im\sigma}\hat {f}_{im\sigma}\rangle_f
\\
-\sum_{ii'mm'\sigma} t_{imi'm'\sigma}  \langle \hat f^\dag_{im\sigma} \hat f_{i'm'\sigma}\rangle_f
\langle \hat O_{i\alpha}^\dag \hat O_{i'\alpha'}\rangle _s\,.
\label{eq:Etotal}
\end{multline}

We now apply this approach to the one-dimensional single band Hubbard model at half filling of Eq.~(\ref{eq:1dHubH}).  Without loss of generality, we choose $b_{i\uparrow}=-b_{i\downarrow}$ to break spin symmetry on each site $i$. For ferromagnetic (FM) order, we choose aligned symmetry breaking fields between neighboring sites $b_{i+1,\sigma}=b_{i\sigma}$, while AFM order requires staggered fields $b_{i+1,\sigma}=-b_{i\sigma}$.  Hence, the field strength $b$ for spin up at one site is sufficient so specify the fields at all sites.  We numerically solve the spinon+slave self-consistent equations using the single-site mean-field approximation described Section~\ref{sec:meanfield}.

\begin{figure}
\includegraphics[width=3in]{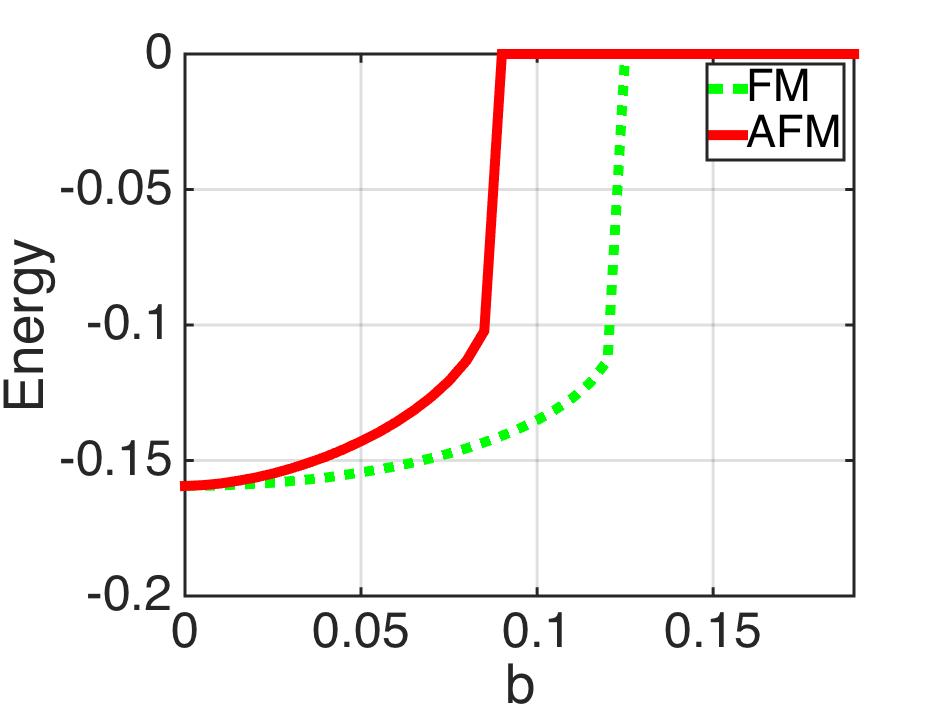}
\includegraphics[width=3in]{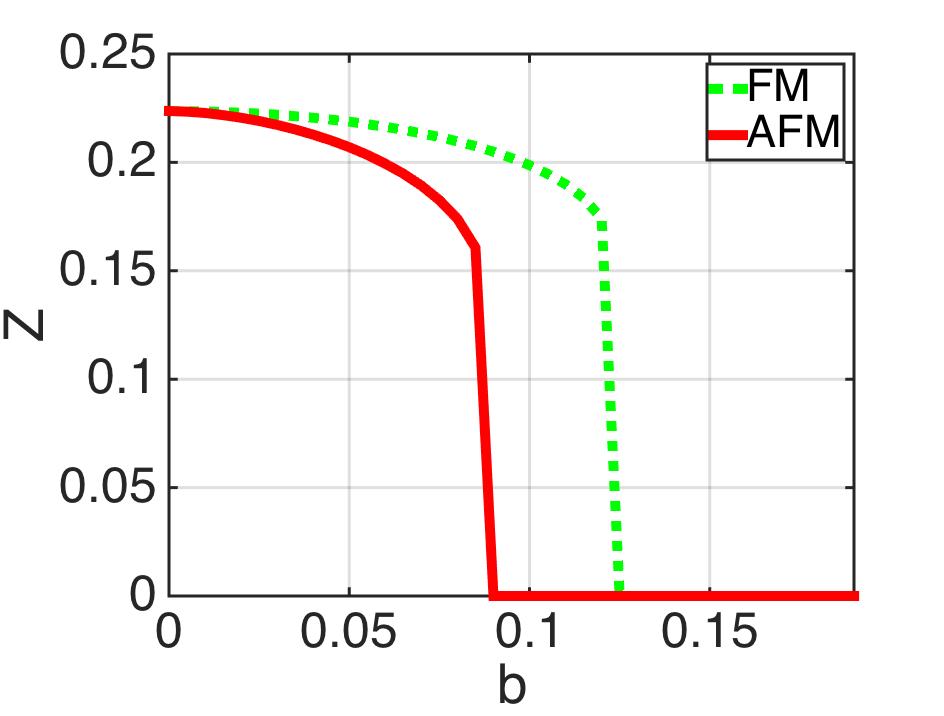}
\caption{\label{fig:EZvsbslaverotor} Total energy per site and quasiparticle weight $Z$ (renormalization factor)versus symmetry breaking perturbation field strength $b$ based on the slave-rotor method for the half-filled single-band 1D Hubbard model with $U=2$ and $t=1$. 
}
\end{figure}

We begin our analysis with the most coarse-grained slave-boson representations that only describe the total electron count on each site (i.e., no information on the spin configuration).  These are the slave-rotor and number-slave methods.  The chief difference between them is that the number count on a site can be any integer in the slave-rotor method while the number-slave corrects this by only permitting the electron count to be among the physically allowed values (e.g., zero, one or two for the single band Hubbard model).  Figure~\ref{fig:EZvsbslaverotor} show the dependence of the total energy and quasiparticle weight $Z$ (i.e., renormalization factor) on the field strength $b$ within the slave-rotor approach.
For the slave-rotor, increasing $b$ increases the total energy of both AFM and FM solutions: the non-magnetic solution is the preferred ground state.  The strength of electronic correlations, measured by how much $Z$ deviates from its non-interacting value of unity, also increases with $b$. This $b$ dependence is opposite to what one would expect for the actual  model system: a more spin-polarized system should have smaller number fluctuations as occupancies are driven towards one or zero and the electron configuration becomes better described by a single Slater determinant.  Finally, the slave-rotor predicts an abrupt transition to a Mott insulator at finite $b$ which is  peculiar (and wrong).

\begin{figure}
\includegraphics[width=3in]{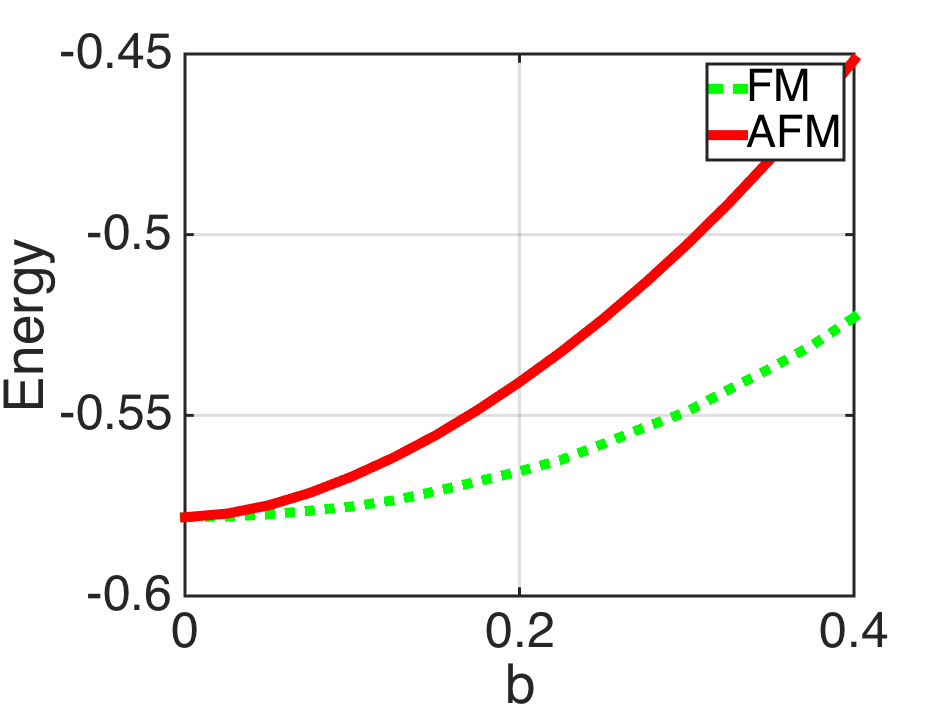}
\includegraphics[width=3in]{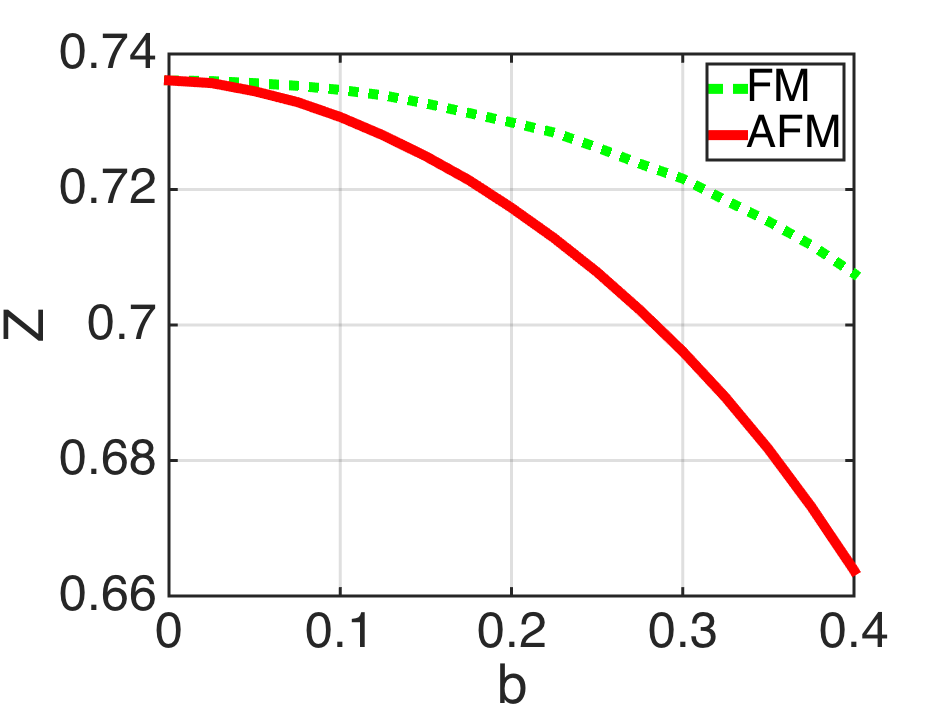}
\caption{\label{fig:EZvsbnumberslave} Total energy per site and $Z$ versus field strength $b$ for the number-slave method for the single-band 1D Hubbard model at half filling with $U=2$ and $t=1$.
}
\end{figure}

The number-slave results for total energy and $Z$ versus $b$, displayed in Figure~\ref{fig:EZvsbnumberslave}, are somewhat of an improvement over those of the slave-rotor but are still fundamentally flawed.  The energy is still minimized by the non-magnetic solution at $b=0$ (although the energy rises more gently with $b$) and $Z$ drops with $b$ (albeit more modestly). The failure of the slave-rotor and number-slave methods is tied to the fact that they do not consider the spin degree of freedom. 

\begin{figure}
\includegraphics[width=3in]{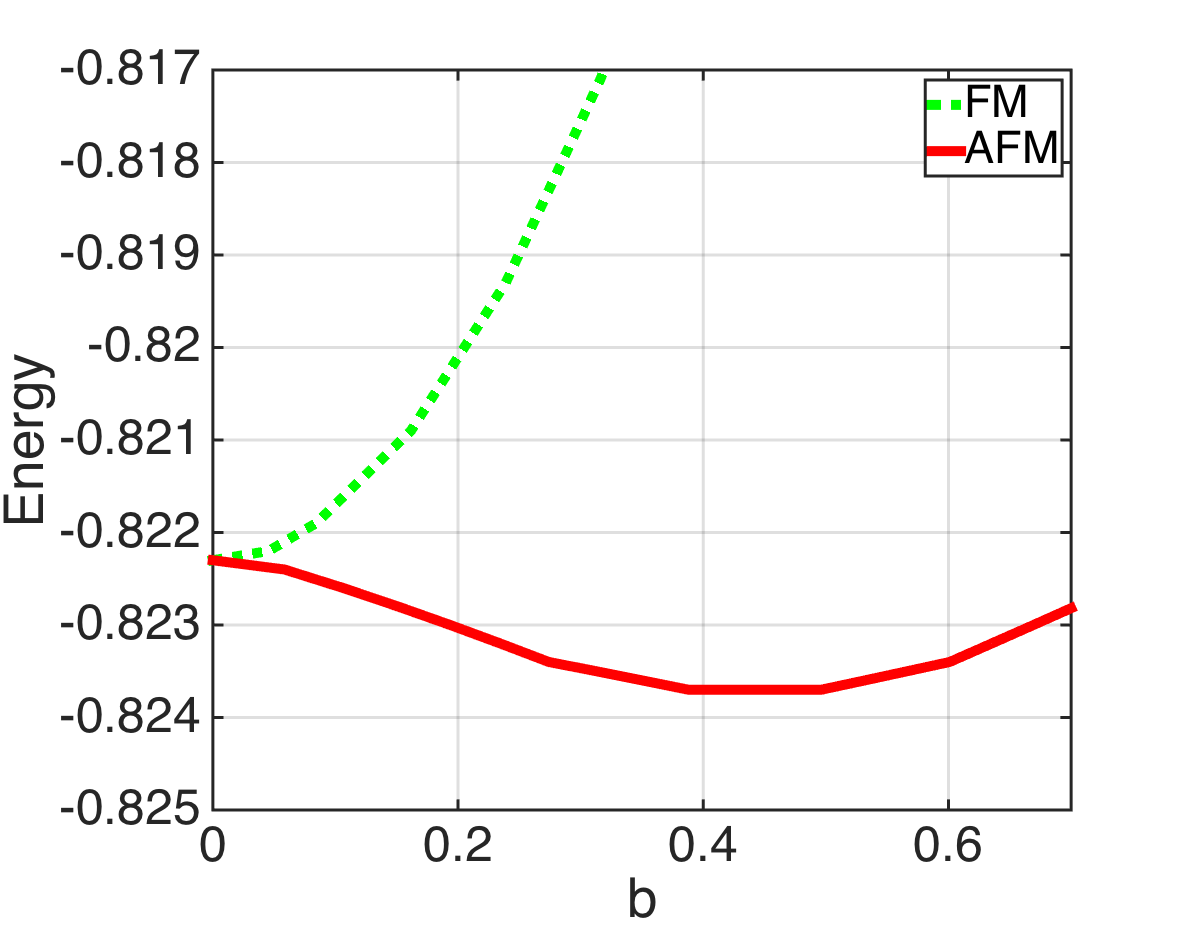}
\includegraphics[width=3in]{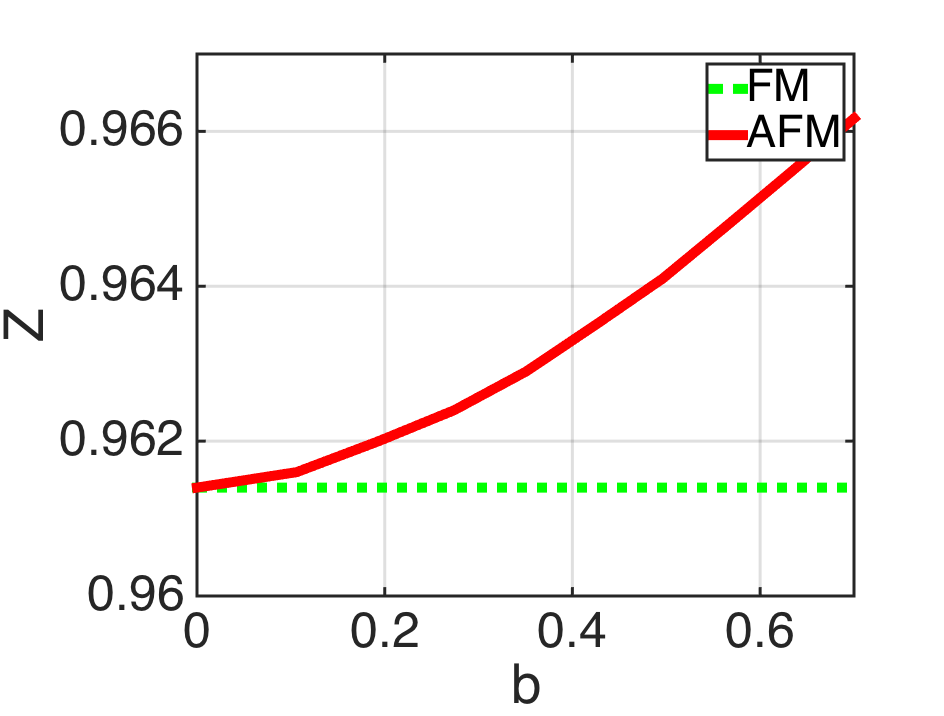}
\caption{\label{fig:EZvsbspinorbslave} Total energy per site and $Z$ versus field $b$ for the spin+orbital-slave approach for the single-band 1D Hubbard model at half filling with $U=2$ and $t=1$. Unlike the number-slave and slave-rotor,  correlations decrease with increasing  $b$ for the AFM phase and slowly increase with b for the FM phase.}
\end{figure}

\begin{figure}
\includegraphics[width=3in]{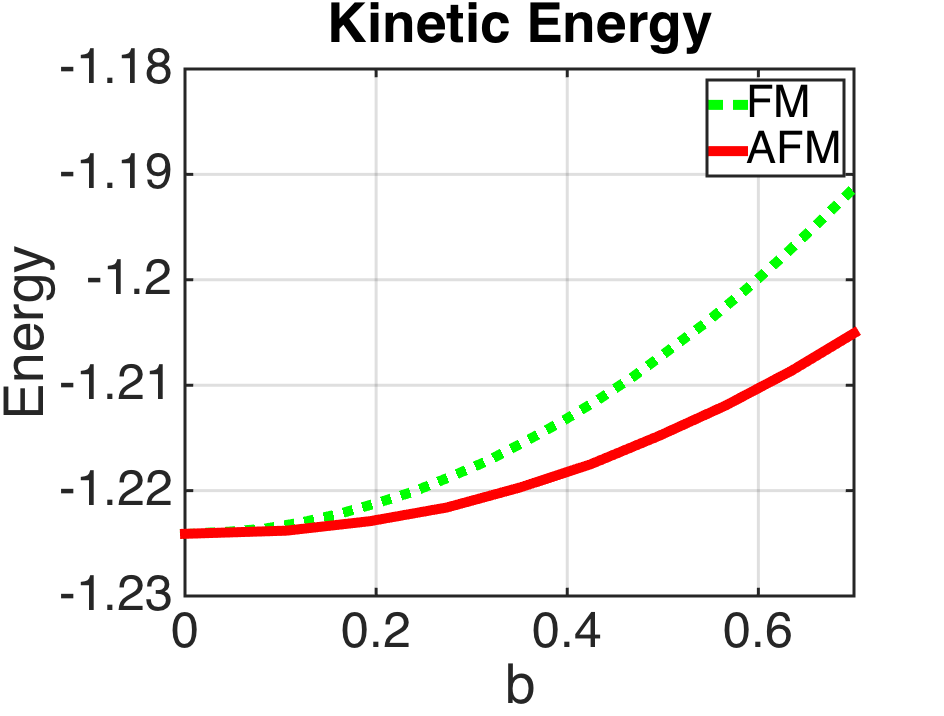}
\includegraphics[width=3in]{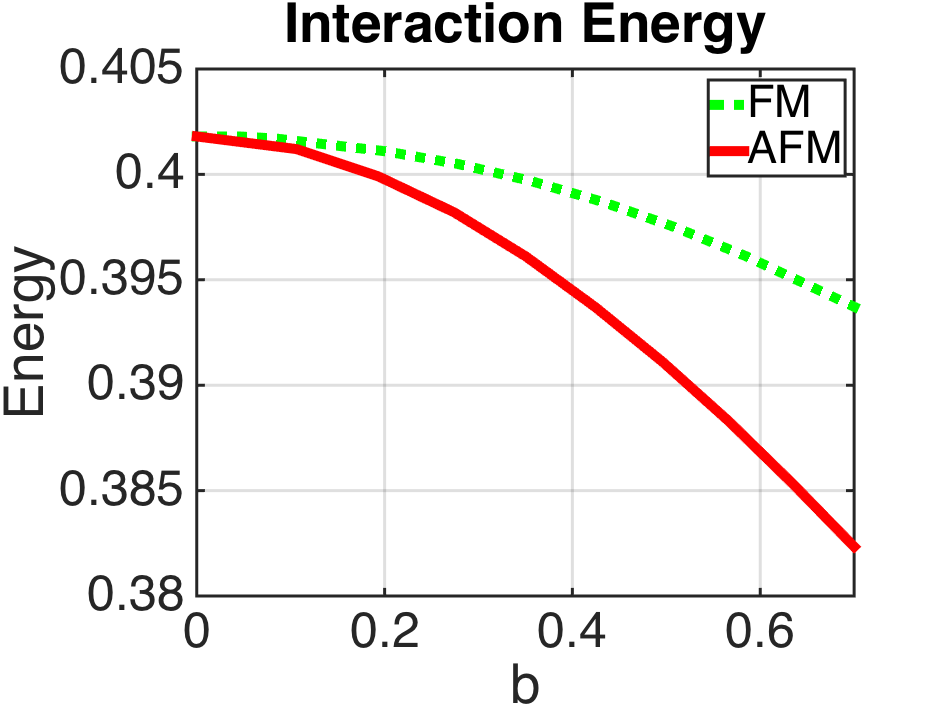}
\caption{\label{fig:KinIntSpin} }
\end{figure}

Due to the simplicity of the single-band Hubbard model, the only remaining slave model is the spin+orbital-slave approach (called ``spin-slave'' in the literature \citep{Hassan2010,DeMedici2005,DeMedici2011}).  On each site, the each spin channel has its own dedicated slave particle.  The energy versus $b$ plot in Figure~\ref{fig:EZvsbspinorbslave} shows that we obtain an AFM ground state since a minimum appears at finite $b$.  The figure also shows that the degree of electronic correlation is reduced with increasing $b$ (and increasing strength of AFM order) as the occupancies get closer to zero and one: the system becomes less strongly interacting as $b$ is strengthened.  This is what we expect: with increasing AFM spin-polarization, the electronic configuration of the system is driven to extremes of occupation (zero or one for each spin channel) meaning that one can describe the system more accurately with a single (non-interacting) Slater determinant.  More details on the energetic behavior versus $b$ is provided by Figure~\ref{fig:KinIntSpin} where the individual components of the total energy are shown versus $b$.  The interaction energy (Hubbard $U$ term) is reduced by the spin symmetry breaking since for both FM and AFM order the occupancies move away from half-filling where occupancy fluctuation is largest.  The band (hopping or kinetic) energy rises with $b$ due to the splitting of bands upon symmetry reduction. Both behaviors are generic and as expected.  However, the  reason the AFM order shows a minimum total energy versus $b$ is due to the fact that $Z$ becomes larger with $b$ in this case: a larger $Z$ (i.e., larger $\langle O\rangle$) will enhance hopping and widen the bands and thus offset the reduction of total band energy due to the creation of spin polarization.

The take-home message of this section is that the introduction of symmetry breaking fields can succeed in stabilizing symmetry-broken ground states due to electronic correlations as long as the slave approach being used is able to describe the symmetry breaking degree of freedom (spin in the 1D single band Hubbard model).  We are thus motivated to improve upon the {\it ad hoc} nature of the approach and put it on a firmer theoretical in the next section.

\section{Self-consistent total energy approach}

In this section, we justify the successful but {\it ad hoc} approach of the previous section. Namely, we describe a total energy functional that can be applied to any type of slave-particle problem and which permits easy incorporation of the various types of desired constraints.  Specifically, we show that the slave-particle approach is a variational approach to the interacting ground-state problem, and we provide an explicit form for the variational energy functional. We also show that this viewpoint provides significant  practical benefits for efficient solution of the self-consistency problem between slave and spinon sectors.

The form of the energy functional $F$ is given by
\[
F = E_{total} + \mbox{constraints}
\]
where $E_{total}$ is from Eq.~(\ref{eq:Etotal}) and the constraint terms are enforced by Langrange multipliers.  

Prior to the introduction of symmetry breaking fields, the constraints we have enforced are that $\langle N_{i\alpha}\rangle_s=\langle \hat n_{i\alpha}\rangle_f$ as well as the normalization of the  spinon and slave wave functions $\braket{\Psi_f}{\Psi_f} = \braket{\Phi_s}{\Phi_s}$.  To incorporate symmetry breaking fields, we choose to parametrize the functional $F$ by {\it target} spinon occupancies $\nu_{im\sigma}$: these numbers are the occupancies that we are constraining the spinons to obey, i.e., the constraints are $\langle n_{im\sigma}\rangle_f = \nu_{im\sigma}$.  The associated Lagrange multipliers are $b_{im\sigma}$.  Hence the energy functional has the form, where we write out $E_{total}$ explicitly,
\begin{multline}
F(\{\nu_{im\sigma}\}) = \sum_i \langle \hat H^i_{int}\rangle_s+\sum_{im\sigma}\epsilon_{im\sigma} \langle\hat {f}^\dag_{im\sigma}\hat {f}_{im\sigma}\rangle_f
\\
-\sum_{ii'mm'\sigma} t_{imi'm'\sigma}  \langle \hat f^\dag_{im\sigma} \hat f_{i'm'\sigma}\rangle_f
\langle \hat O_{i\alpha}^\dag \hat O_{i'\alpha'}\rangle _s \\
- \lambda_f[\braket{\Psi_f}{\Psi_f}-1] - \lambda_s [\braket{\Phi_s}{\Phi_s} -1]\\
- \sum_{i\alpha} h_{i\alpha}[ \expect{\hat n_{i\alpha}}_f - \expect{\hat N_{i\alpha}}_s]\\
- \sum_{im\sigma} b_{im\sigma}[\expect{\hat n_{im\sigma}}_f - \nu_{im\sigma}]\,.
\label{eq:Falmost}
\end{multline}
The Lagrange multiplies $\lambda_f$ and $\lambda_s$ enforce normalization of the spinon and slave wave functions, respectively.  The $h_{i\alpha}$ enforce particle number matching between slave and spinon sectors.  The $b_{im\sigma}$ enforce spinon particle matching to target values. As expected, when the constraints are obeyed, $F=E_{total}$.
 
The point of having a energy functional is that the minimizing variational conditions, which generate  desired eigenvalue problems, are easily derived by differentiation. In addition, the value of $F$ provides a variational estimate of the ground state energy. Setting the derivative versus $\bra{\Psi_f}$ to zero gives the spinon eigvenalue equation
\[
0 = \frac{\delta F}{\delta \bra{\Psi_f}} = H_f\ket{\Psi_f} - \lambda_f\ket{\Psi_f}
\]
where the spinon Hamiltonian is that of Eq.~(\ref{eq:Hspinonwithfield}) which includes the symmetry breaking fields.  Similarly, the minimum condition for $\ket{\Phi_s}$ gives a slave eigenvalue problem with the slave Hamiltonian of Eq.~(\ref{eq:Hslave}).  

The above formalism shows that, once all the constraints are obeyed, $F(\{\nu_{im\sigma}\})=E_{total}(\{\nu_{im\sigma}\})$.  The remaining task it to search over the target occupancies $\nu_{im\sigma}$ to find the minimum total energy.  While theoretically straightforward, in practice such an approach is difficult and inefficient because for each specified $\{\nu_{im\sigma}\}$, one must find the fields $b_{im\sigma}$ that enforce those particular target occupancies: this requires solving the spinon+slave problem a great many times.  

Practically, it is better to use the $b_{im\sigma}$ as the independent variables and to minimize the energy over the (formally, this corresponds to a Legendre transformation of $F$).  Hence, we now view $\nu_{im\sigma}$ as whatever mean spinon occupancies are generated by solution of the spinon+slave problem at fixed $\{b_{im\sigma}\}$ which makes that corresponding constraint form always vanish.  Hence, in what follows, we will use the symmetry breaking fields as independent variables and consider the total energy functional $F(\{b_{im\sigma}\})$.  Since we will always be obeying the key constraints for a physical solution,  $F(\{b_{im\sigma}\})=E_{total}(\{b_{im\sigma}\})$ will be true.  Hence, minimization of the total energy versus $\{b_{im\sigma}\}$ will coincide with minimization of $F$.

\section{Simplified and more efficient slave-particle approach}

Up to this point, the slave-particle approaches we have developed require self-consistency between spinon and slave sectors in a specific manner: not only do the spinon expectations renormalize slave hopping terms (and conversely for slave expectations and spinon hoppings), but a shared set of Lagrange multipliers $h_{i\alpha}$ enforce particle number matching $\expect{\hat n_{i\alpha}}_f = \expect{\hat N_{i\alpha}}_s$.  The process of finding the $h_{i\alpha}$ is numerically challenging: the $h_{i\alpha}$ appear with opposite signs in the spinon $H_f$ and slave $H_s$ Hamiltonians meaning that increasing $h_{i\alpha}$ decreases $\expect{\hat n_{i\alpha}}_f$ but increases $\expect{\hat N_{i\alpha}}_s$.  Our general observation is that this ``fighting'' over $h_{i\alpha}$ between the slave and spinon sectors leads to a time-consuming self-consistent process requiring many iterations to reach convergence.   

Accelerating this process requires a simple change of variables that is motivated by three related observations:
 (i) in the total energy functional of Eq.~(\ref{eq:Falmost}), the spinon and slave number constraints are not treated symmetrically because the spinons have the added $b_{im\sigma}$ terms, (ii) in the spinon Hamiltonian of Eq.~(\ref{eq:Hspinonwithfield}), we can add the $h_{i\alpha}$ and $b_{im\sigma}$ terms together into a single term whereas the slave Hamiltonian of Eq.~(\ref{eq:Hslave}) only has the $h_{i\alpha}$ terms, and (iii) in the end, these Lagrange multipliers $h_{i\alpha}$ and $b_{im\sigma}$ do not appear in the total energy so  rearranging them in various ways does not change the total energy. 

For the spinon Hamiltonian, we consider instead the new symmetry breaking field  given by the  sum $B_{im\sigma} = h_{i\alpha} + b_{im\sigma}$.  The spinon Hamiltonian is now
\begin{multline}
\hat{H}_f=\sum_{im\sigma}\epsilon_{im\sigma}\hat{f}_{im\sigma}^\dag\hat{f}_{im\sigma}
-\sum_{im\sigma}B_{im\sigma}\hat{f}_{im\sigma}^\dag\hat{f}_{im\sigma}\\
-\sum_{ii'\alpha\alpha'} \langle\hat{O}^\dag_{i\alpha}\hat{O}_{i'\alpha'}\rangle_s\!
\sum_{\substack{m\sigma\in\alpha\\m'\sigma\in\alpha'}}
t_{imi'm'\sigma}  \hat{f}^\dag_{im\sigma}\hat{f}_{i'm'\sigma}\,
\label{eq:HspinonbigB}
\end{multline}
while the slave Hamiltonian is unchanged
\begin{multline*}
\hat{H}_{s}=\sum_i \hat H^i_{int} +\sum_{\alpha}h_{i\alpha}\hat{N}_{i\alpha}\\
-\sum_{ii'\alpha\alpha'}\left[
\sum_{\substack{m\sigma\in\alpha\\m'\sigma\in\alpha'}}t_{imi'm'\sigma}\langle \hat{f}^\dag_{im\sigma}\hat{f}_{i'm'\sigma} \rangle_f \right]
 \hat{O}^\dag_{i\alpha}\hat{O}_{i'\alpha'}\,.
\end{multline*}
The slave Hamiltonian $H_s$ no longer shares a common Lagrange multiplier with the spinon Hamiltonian $H_f$.

Operationally, this means that when we solve the slave Hamiltonian problem, we are given specified $\expect{\hat n_{i\alpha}}_f$ as input, and we solve the slave  problem 
while adjusting the $h_{i\alpha}$ so as to ensure that the  slave-particle counts match the input: $\expect{\hat N_{i\alpha}}_s = \expect{\hat n_{i\alpha}}_f$.  However, when solving the spinon problem in the presence of symmetry breaking fields $B_{im\sigma}$, there is no need to do particle number matching: the Lagrange multiplier $B_{im\sigma}$ simply make the spinon particle counts match some free floating  values.  In this way,  particle number matching between the slave and spinon sector is decoupled which {\it grealy} simplifies the   self-consistency process.  Put another way, the symmetry breaking fields $B_{im\sigma}$ specify a set of desired spinon particle counts $\{\nu_{im\sigma}\}$, and  the slave sector is required to match this particle numbers via the $h_{i\alpha}$ Lagrange multipliers.

We find that this simplified approach, which is  equivalent to the  standard approach of having $h_{i\alpha}$ appear in both Hamiltonians, is much more efficient in numerical calculations as it greatly speeds up self-consistency.    In this new approach, one achieves rapid self-consistency for a given set of $\{B_{im\sigma}\}$ which specify the spinon Hamiltonian and the target spinon occupancies $\nu_{im\sigma}$.  One can then minimize $E_{total}(\{B_{im\sigma}\})$ over the $B_{im\sigma}$ to find the symmetry-broken ground state.  In our experience, this new approach requires $\sim$5-10 times fewer self-consistent steps to reach the same level convergence.

\begin{figure}
\includegraphics[scale=0.25]{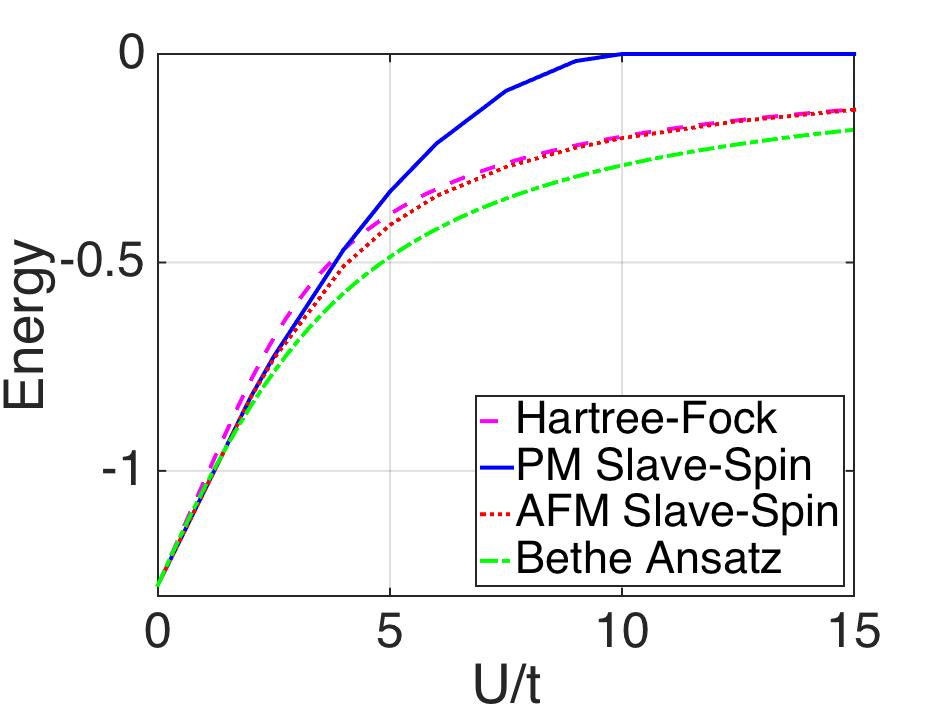}
\caption{\label{fig:PMAFMBetheEtot} Comparison of the ground state energies (in units of $t$) for the single-band 1D Hubbard model at half filling based on the AFM Hartree-Fock solution, the PM slave-spin solution, the symmetry broken (AFM) slave-spin ground state solution, and the exact Bethe Ansatz (AFM) solution as calculated by the method of Ref.~\onlinecite{Sanchez1998}.}
\end{figure}
Using this  method, we can rapidly scan over $B$ in a stable, self-consistent way to obtain ground state energies. Figure~\ref{fig:PMAFMBetheEtot} shows the dependence of the ground state energy of the half-filled single-band 1D Hubbard model as a function of $U/t$: for each $U/t$, we easily scan over the new symmetry breaking field strength $B$ to find the AFM ground state energy.  The figure shows energy versus $U/t$ for the AFM state as well as the $B=0$ non-magnetic solution compared to the exact Bethe ansatz solution for this problem.\citep{Lieba2009}  Overall, the comparison between the AFM slave-spin  solution (which is insulating in the spinon sector) and the exact Bethe ansatz is satisfactory given the simplicity of the single-site mean field slave model used here.  As expected, the AFM slave-spin method becomes very much like AFM Hartree-Fock in the large $U/t$ limit of very strong spin polarization since both approaches essentially describe the system as a single Slater determinant.  We note that the non-magnetic ground state has an incorrect evolution from a metallic system at small $U/t$ to a Mott-insulating phase at $U/t\ge 10$.

\section{Conclusion}
\label{sec:conclusion}

We've shown how slave particle methods can be used to obtain spontaneously symmetry-broken electronic phases based on a total-energy approach.  We have described and tested our ideas on the classic 1D Hubbard model Hamiltonian. Furthermore, we have shown how to enable symmetry breaking via the use of auxiliary symmetry breaking fields in a self-consistent way that greatly lowers the computational burden and stability from the standard slave-particle calculation. Further, we have demonstrated that in order to obtain spontaneously symmetry-broken phases in the spinon sector, the slave-sector must be allowed to break the corresponding symmetry explicitly by having different slave-modes for the different degrees of freedom which may undergo symmetry breaking.

\section{Acknowledgements}

This work was supported  by the National Science Foundation via Grant MRSEC DMR 1119826.

\bibliography{Mendeley}

\end{document}